\documentclass[11pt]{article} 
\usepackage{hyperref} 
\pdfoutput=1 
 
\begin{document} 
 
\title{The Clapping book} 
 
\author{Pedro M. Reis and John W. M. Bush \\ 
\\\vspace{6pt} Department of Mathematics, \\ Massachusetts Institute of Technology, \\ 77 Massachusetts Avenue Cambridge, MA 02138, USA} 
 
\maketitle

\begin{abstract} 

A steady horizontal air stream flows across a book \cite{dyke,GFM} clamped at its downstream end. Pages lift off to form a growing bent stack whose shape is determined by the torques associated  with aerodynamic forces, weight and elastic resistance to  bending.  As
more pages lift off to join the bent stack, the increasing importance of bending rigidity to dynamic pressure eventually causes the book to clap shut. The process restarts, and self-sustained  oscillations emerge. [Fluid dynamics video] 

\end{abstract} 

\section{Introduction} 
Controlled experiments are performed in a wind tunnel with cross-sectional area of $(0.3\times0.3)m^2$ that we operate at wind speeds in the range $(1<U<10)m/s$. The stack consists of 200 pages of paper sheets (thickness $t=76\mu m$, density $\rho=9842g/m^3$, bending stiffness $B=8.55\times10^{-3}Nm$) with dimensions $(0.029\times0.17)m^2$. In this regime, extremely regular oscillations are attained with frequencies in the range of $(0.1<f<1)Hz$.

The videos associated with this submission for the 26th Annual Gallery of Fluid Motion can be found in \href{http://ecommons.library.cornell.edu/bitstream/1813/11473/2/GFM_clapping_reis_mpeg1.mpg}{ low (mpeg-1)} and \href{http://ecommons.library.cornell.edu/bitstream/1813/11473/3/GFM_clapping_reis_mpeg2.mpg}{ high (mpeg-2)}  resolution.

\end{document}